\documentclass{article}
\usepackage[english]{babel}
\usepackage{csquotes}
\usepackage{authblk}

  \usepackage[
    backend=biber,
    style=nature,
    sorting=none,
  ]{biblatex}

\usepackage[letterpaper,top=2cm,bottom=2cm,left=3cm,right=3cm,marginparwidth=1.75cm]{geometry}

\usepackage{amsmath}
\usepackage{graphicx}
\usepackage{gensymb}
\usepackage[colorlinks=true, allcolors=blue]{hyperref}
\usepackage{booktabs}

\newcommand{\coo}{\ensuremath{\mathrm{CO_2}} }
\newcommand{\meth}{\ensuremath{\mathrm{CH_4}} }
\newcommand{\nit}{\ensuremath{\mathrm{N_2O}} }
\newcommand{\NPP}{\ensuremath{\mathrm{NPP}} }
\newcommand{\RH}{\ensuremath{\mathrm{RH}} }

\author[1,a*]{Hannah B\"{a}ck}
\author[2,a*]{Riley May}
\author[3]{Divya Sree Naidu}
\author[3]{Steffen Eikenberry}
\affil[1]{Corresponding Author, University of Iowa}
\affil[2]{Utah State University}
\affil[3]{Arizona State University}
\affil[a]{Simon A. Levin Mathematical, Computational, and Modeling Sciences Center: Quantitative Research for the Life \& Social Sciences Program, Arizona State University}
\affil[*]{Equally contributing authors}

\title{Effect of Methane Mitigation on Global Temperature under a Permafrost Feedback}

\begin{document}
\maketitle

\subsection*{Acknowledgements}
We would like to thank Dr. Fabio Milner, Director of the Simon A. Levin Mathematical, Computational and Modeling Sciences Center (Levin Center), for giving us the opportunity to participate in the Quantitative Research in the Life and Social Sciences program. We would also like to thank Co-Directors Dr. Abba Gumel and Dr. John Nagy for their efforts in planning and executing the program’s instruction and activities. We also recognize the work of the many administrative staff and tutors who supported this effort.  This research was conducted as part of 2022 QRLSSP at the Levin Center (MCMSC) at Arizona State University (ASU). This project has been partially supported by grants from the National Science Foundation (NSF Grant-DMS-1757968 and NSF Grant FAIN-2150492), the National Security Agency (NSA Grant H98230-20-1-0164), the Office of the President of ASU, and the Office of the Provost of ASU.

\newpage

\subsection*{keywords}
climate modeling, global warming, emissions, methane, mitigation, permafrost

\begin{abstract}

Earth systems may fall into an undesirable system state if 1.5\degree C of warming is exceeded.

Carbon release from substantial permafrost stocks vulnerable to near-term warming represents a positive climate feedback that may increase the risk of 1.5\degree C warming or greater. Methane (CH$_4$) is a short-lived but powerful greenhouse gas with a global warming potential 28.5 times that of carbon dioxide (CO$_2$) over a 100 year time span.  Because permafrost thaw in the coming centuries is partly determined by the warming in the 21st century, rapid reductions in methane emissions early in the 21st century could have far reaching effects. We use a reduced complexity carbon cycle model and a permafrost feedback module to explore the possibility that accelerating reductions in methane emissions could help avoid long-term warming by limiting permafrost melt.  We simulate 3 extended Representative Concentration Pathway (RCP) emission scenarios (RCP 2.6, 4.5, and 6) through the year 2300 and impose methane mitigation strategies where we reduce \meth emissions by 1\%, 5\% or 10\% annually until the long-term scenario emission level is reached. We find that accelerated rates of methane mitigation do not sufficiently alter the global temperature anomaly to prevent or delay a permafrost feedback, nor do they result in meaningful long term reductions in temperatures. We find that the long-term magnitude of methane mitigation (i.e., long-term emission level) and not the rate of reduction, corresponds to long-term temperature change. Therefore, policy and mitigation efforts should emphasize durable decreases in methane emissions over rapidity of implementation.

\end{abstract}

\newpage
\section{Introduction}
There is increasing recognition and concern that multiple intrinsic positive feedbacks in the Earth system may lead to ``tipping points’’ whereby the Earth system state may rapidly transition into a new undesirable one once some threshold of global mean warming is passed. Such feedbacks include ice sheet loss, permafrost melt, Atlantic ocean circulation, and Amazon rainforest dieback \cite{lenton2019}. Without these feedbacks, previous estimates for ``safe warming’’ ranged between 2-3\degree C for ``safe warming." With the inclusion of positive, interacting feedbacks, Lenton et al. \cite{lenton2019} concluded that warming must be limited to 1.5\degree C. Warming past 1.5\degree C may cause runaway feedback effects. Therefore, a full assessment of mitigation pathways to avoid crossing undesirable climate boundaries should explicitly include major feedbacks.

Among the many positive climate feedbacks, permafrost melt and subsequent carbon emissions may have the greatest effect on climate in the coming centuries due to the vast amounts of carbon stored within the permafrost and vulnerability to  temperature perturbation (\cite{crichton2016}). Because long-term warming is primarily a function of \textit{cumulative} carbon emissions \cite{ph2014} \cite{ipcc15degrees} the concept of the carbon budget was developed. It describes a total cumulative amount of carbon that can be emitted to limit global mean warming (with some degree of confidence) to a given level, e.g. 1.5\degree C. Each year, anthropogenic \coo emissions take roughly 9 GtC off the budget. In 2019, there was an estimated 500 GtC left in the budget for 1.5\degree C, \cite{lenton2019} \cite{Rogelj2019} but more recent estimates from 2021 posit that only 420 GtC remain \cite{FPcarbonbudget2021}.

The permafrost may be the most impactful near-term feedback because of high net carbon content and sensitivity to degradation with increased warming. As temperatures rise, permafrost emissions could rapidly exhaust the carbon budget. As of 2014, estimates concluded that the circumpolar permafrost at a depth of 0-3 meters stored 1035 gigatonnes of carbon (GtC) \cite{hugelius2014}. The Intergovernmental Panel on Climate Change (IPCC) has created emission scenarios, representative concentration pathways (RCPs), namely RCP2.6, RCP4.5, RCP6.0, and RCP8.5, with the number corresponding to the change in radiative forcing the Earth experiences in the year 2100 (e.g. under RCP 8.5, the Earth will experience +8.5 watts/m$^{-2}$ of radiative forcing). Under RCP 8.5, 33-114 GtC may be released from the permafrost by 2100, contributing to an additional warming of 0.04-0.23\degree C \cite{deimling2012}, and by 2300, half the vulnerable permafrost carbon stock could be released. While the RCP 8.5 is a high-end scenario, there is still a great deal of uncertainty regarding permafrost carbon stocks, thawing processes, and subsequent microbial decomposition of \coo and \meth, as well as other potential feedbacks \cite{deimling2012}. Another study found that the permafrost feedback effect reduces the carbon budget, when set at avoiding 2\degree C, by 100 GtC \cite{gasser2018}, accounting for roughly 24\% of our remaining carbon budget \cite{FPcarbonbudget2021}. Therefore, permafrost feedback is the primary positive climate feedback we examine in this paper.
Methane is a powerful greenhouse gas and is responsible for about 1.19 W/m$^{-2}$ of radiative forcing since 1750. Given that warming is a function of multiple anthropogenic greenhouse gas emissions, mitigating methane emissions will play an important role in avoiding the climate boundary \cite{dreyfus2022} \cite{rogelj2014} \cite{harmsen2019}  \cite{xu2017} \cite{christensen019}. However, multiple authors have argued that methane mitigation is not a substitute for immediate decarbonization and will not ``buy us time’’ as we work towards net-zero carbon \cite{ph2014} \cite{rogelj2014} \cite{bowerman2013} \cite{myhre2011}. These conclusions hinge on the fundamentally different climate effects of short-lived and long-lived greenhouse gases. 

Because near term climate forcers like methane have relatively short atmospheric lifetimes, the annual emissions \textit{rate} of these substances is far more relevant than cumulative emissions over time \cite{ph2014}: There is no long-term ``methane budget''. We are concerned with net cumulative emissions of \coo because it is a long-lived millennial gas, where in contrast, methane oxidizes to \coo in the presence of hydroxyl radicals (OH), and has a perturbation lifetime of approximately 12.4 years \cite{ar6ch6}. Therefore, mitigation efforts should make their focus limiting annual methane emissions rather than limiting a net cumulative amount.
 
How to best plan methane mitigation in the coming years remains an open question. Although there is uncertainty regarding how much, there is general agreement that some warming can be avoided in this century by implementing methane mitigation plans \cite{smith2013} \cite{dreyfus2022}, with methane mitigation especially valuable for limiting near term warming rates and magnitude by mid-century \cite{ocko2021}. However, Mckeough et al. \cite{mckeough2022} concluded that mitigation could be postponed to as late as 2050-2080 with limited impact to temperature anomaly at 2100 \cite{mckeough2022}. Nevertheless, methane mitigation is understudied in relation to its long-term effects (past 2100), especially in relation to positive climate feedbacks. That is, because feedbacks such as permafrost may be sensitive to near-term temperature changes that are strongly influenced by methane emissions rates early in the twenty-first century, rapid methane mitigation could be a higher priority than previously realized. 
 
We address the question of the rate of methane mitigation: Is there any benefit to an accelerated emissions reduction plan in the presence of a permafrost carbon feedback? We are interested in the possibility that the mitigation of methane could prevent some release of \coo, as well as \meth, from the permafrost, and therefore mitigate long-term warming. Because methane has considerable short term temperature effects and can strongly influence the short-term rate of warming, it warrants examination in relation to the permafrost feedback. 
 
We use a reduced complexity model that represents the most relevant Earth systems to understand the dynamics between reducing annual \meth emissions and long term temperature changes from \coo and \meth, as well as nitrous oxide (\nit), due to its interaction with methane. Radiative forcing attributable to our three greenhouse gases is described with IPCC formulas and a two-box ocean layer heat transfer model due to Pierrehumbert \cite{ph2014}. We develop a simplified carbon cycle model based on Glotter et al. (2014) and Hartin et al. (2015, 2016) that describes flux between three terrestrial biota compartments and two ocean compartments. Increased carbon uptake due to \coo fertilization and \coo buffering with acidification are included. We model permafrost decomposition via a linear decrease in permafrost extent with rise in temperature and subsequent release of emissions as exponential decay. This is based on the model of Kessler et al. \cite{kessler2017}, where permafrost carbon mobilization to a labile carbon pool increases linearly with temperature. Carbon in this labile pool is then emitted to the atmosphere as a mix of \coo and \meth according to exponential dynamics with an e-folding time of about 70 years. These simplified representations of Earth systems describe the relationship of emissions to temperature perturbation. 

We then drive the model using RCP \coo emission time-series while varying rates of methane mitigation to determine the impact on long term temperature. We find that the magnitude of methane mitigation (i.e., the final long-term sustained methane emission rate that is attained) has a relationship to long term temperature, but not the rate of mitigation (i.e, how quickly this final rate is able to be attained). Moreover, these conclusions are \textit{insensitive} to the presence of the permafrost feedback, as well as the time-scale of this feedback (i.e. fast vs. slow carbon decay). Therefore, we conclude that while long-term methane mitigation is essential to climate stabilization goals, early mitigation of net \coo should take priority over accelerated methane mitigation.

\newpage
\section{Methods}
We use a reduced complexity climate model implemented as a system of coupled first order differential equations. These equations model the carbon cycle with carbon fluxes between atmospheric, terrestrial, and ocean layers, along with ocean carbonate chemistry. We use a simple two-box temperature response model to calculate the effects of radiative forcing from \coo, \nit, and \meth on surface temperature. We include a two-part linear permafrost feedback using a differential equation that responds to increases of the global mean temperature. The temperature and carbon cycle model can be driven either by imposed atmospheric GHG concentrations (when driving the model with historical data) or with GHG emissions (and resulting modeled concentrations) when projecting the model forward.

\begin{figure}[h!]
\textbf{Simplified Model Dynamics}\par\medskip
\includegraphics[width=\textwidth]{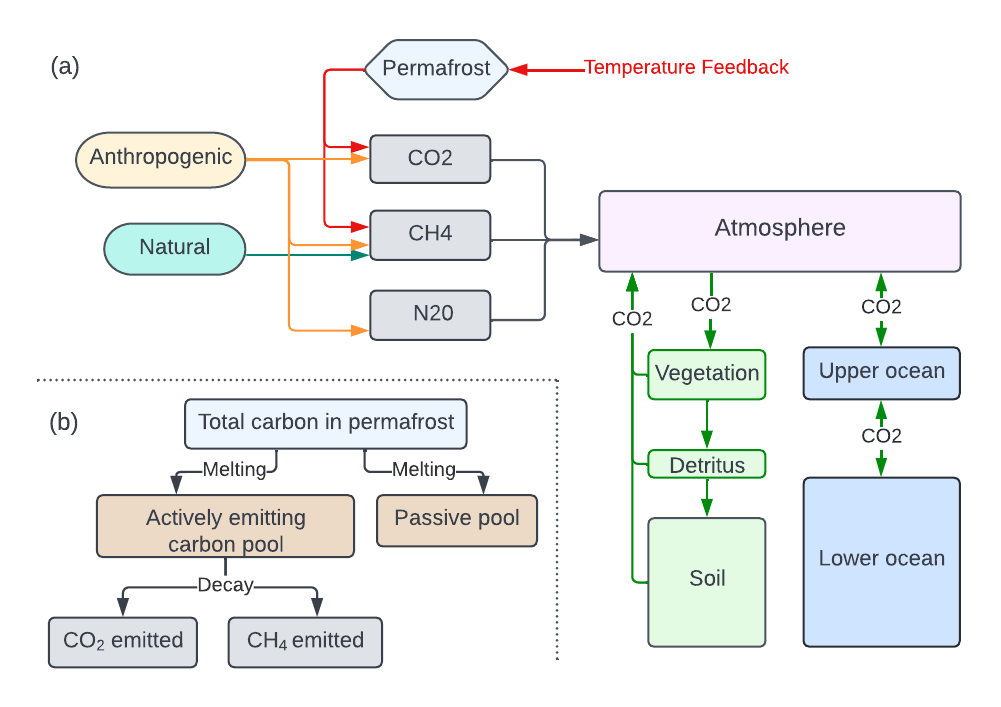}
\caption{Model outline: (a) The model includes \coo, \meth, and \nit emissions into the atmosphere. Note, methane emissions are from anthropogenic and biogenic (natural) sources, and from permafrost emissions. \coo  is from anthropogenic sources and the permafrost. \nit is only from anthropogenic emissions. Carbon cycles between the atmosphere (as \coo),  terrestrial boxes (the vegetation, detritus, and soil), and the ocean boxes (upper and lower). Increases in temperature amplify permafrost emissions. (b) Permafrost sub-model schematic. Total carbon in the permafrost ($C_{PF}$, see Equation \eqref{cpf}) melts to expose soil. In the exposed soil, there is an actively emitting labile carbon pool ($L_C$ and $L_M$, see Equations \eqref{lc} and \eqref{lm}), and a passive or non-labile carbon pool. From the actively emitting carbon pool, carbon is released as  \coo or \meth in constant proportion.}
\label{Fig. 1}
\end{figure}

Figure \ref{Fig. 1} provides a simplified representation of our model's dynamics. Anthropogenic emissions, from fossil fuels and land use change, and the permafrost feedback, are the primary sources of \coo and \meth emissions, our primary warming agents.  Biogenic methane emissions are included as a baseline, non-mitigable \meth source that contributes to warming, set at a constant 300 Mt \meth per year, between the 220-368 Mt estimated by Jackson et al. \cite{Jackson2020}. \coo circulates between the terrestrial and ocean compartments.  Temperature perturbation is calculated from the change in radiative forcing from emissions which warm the upper and, in turn, the lower ocean. This temperature increase then feeds back into the permafrost module, causing permafrost melt and subsequent release of methane and \coo.

\subsection{The Carbon Cycle}
\subsection*{Terrestrial Carbon Stores}
The terrestrial carbon stores included in our model are the vegetation, detritus and soil. Together, these have a significant interaction with atmospheric carbon. We use a differential equation to describe each major box, where $C_V$ is the carbon of the vegetation layer, $C_D$ is the carbon in the detritus layer, and $C_S$ is the carbon in the soil layer. Each box (Equations \eqref{vegeq}, \eqref{detriuseq}, \eqref{soileq}) has a source term representing net primary productivity (NPP), where autotrophs store some of amount of carbon. Each layer has loss and source terms to describe the transfer of carbon from one layer to another. The vegetation box's loss term represents the transfer of carbon to the detritus ($C_V\emph{f}_{vd}$) and soil ($C_V\emph{f}_{vs}$) layers. The detritus box's loss term describes the carbon transfer to the soil layer ($C_D\emph{f}_{ds}$). The soil and detritus boxes have loss terms for heterotrophic respiration (RH), where microbial activity releases carbon into the atmosphere. Finally, each layer has a term representing losses of carbon due to land use change ($F_{LC}$).

We use formulas from Hartin et al. \cite{hector2015} given by 
\begin{align}
    & \frac{dC_V}{dt} = \NPP \emph{f}_{nv} - C_V(\emph{f}_{vd} + \emph{f}_{vs}) - F_{LC}\emph{f}_{lv} \label{vegeq} \text{ ,} \\
    & \frac{dC_D}{dt} = \NPP\emph{f}_{nd} + C_V \emph{f}_{vd} - C_D \emph{f}_{ds} - \RH_{det} - F_{LC} \emph{f}_{ld} \label{detriuseq} \text{ ,}\\
    & \frac{dC_S}{dt} = \NPP\emph{f}_{ns} + C_V \emph{f}_{vs} + C_D  \emph{f}_{ds} - \RH_{soil} - F_{LC} f_{ls} \label{soileq}\text{ .}
\end{align}

Any land use change alters the \coo content of the terrestrial stores. For example, if land is deforested, each layer will release some carbon into the atmosphere. If $x$ GtC are released due to land use change, a certain proportion of those emissions comes from each of the different layer represented by the portions $f_{lv}, f_{ld}, \text{and} f_{ls}$ (Table  \ref{tab:params}). 

Heterotrophic respiration (RH) from the detritus and soil layers feeds into the atmosphere. RH is a function of the current carbon content of the respective strata and the respiration factor $Q_{10}$ which is the factor by which respiration increases for a 10 degree increase in temperature. For soil and detritus, we use a ten-year average of global mean temperature perturbation divided by ten. A running average is used to represent the slow change in temperature from the surface of the Earth to the lower layers of soil.   

Net primary productivity (NPP) is a function of current atmospheric carbon levels compared to pre-industrial carbon levels in ppm scaled by the carbon fertilization parameter, $\beta$. The value for $\beta$ varies depending on the region \cite{hartin2016}, but we use a single value to represent the whole earth system. The total NPP is subtracted from our atmospheric carbon, reflecting the carbon uptake by all terrestrial sinks, and is added to terrestrial stores by a factor $f_{nv}, f_{nd}, f_{ns}$ (table \ref{tab:params}).  Therefore, we determine NPP and RH by
\begin{align}
    & \NPP (t) = \NPP_0 \times f(C_{atm},  \beta) \\
    & f(C_{atm}, \beta) = 1 + \beta \times log \left( {\frac{C_{atm}}{C_0}} \right) \\
    & \RH_{s,d}(t) = C_{s,d} \times  \emph{f}_{rs,rd} \times Q_{10}^{T(t)/10}. 
\end{align}

\newpage
\subsection{Oceanic Carbon Stores}

Ocean carbon flows and buffering chemistry are modeled using the Bolin and Erikson Adjusted Model (BEAM) \cite{glotter2014simple}, where \coo cycles between the atmosphere, upper ocean, and lower ocean. 

We represent the carbon in the atmosphere, upper ocean, and lower ocean boxes as $C_A$, $C_U$, and $C_L$ respectively. 
Important parameters include $\delta_d$ and $\delta_a$, where $\delta_d$ represents the ratio of the moles of the lower (deep) ocean to the moles of the upper ocean. Then, $\delta_a$ is the ratio of moles between the atmosphere and the upper ocean. Given that we know the ocean is approximately 500 times the size of the ocean, we set $\delta_d$ to 50 and solve for $\delta_a$ using known values for atmospheric and ocean moles. We use $k_a$ for the turnover time for the upper ocean, $k_d$ for the turnover time for the deep ocean, and $k_H$ for Henry's constant. Lastly, the anthropogenic emissions stream $E(t)$ is added directly to $C_A$. 

To model the carbon chemistry of the ocean, we look at the dynamics of how atmospheric carbon is absorbed into the different ocean layers. The loss of carbon in the atmosphere to the upper ocean is represented by $-k_a C_A$, and its flux back into the atmosphere is described with the expression $k_a \frac{k_H}{\delta_a}C_U$.  Carbon flux in the upper ocean is described with the same expressions with opposite signs. Additionally, there is an exchange between the upper and lower ocean given by $-k_d C_U + \frac{k_d}{\delta_d} C_L$. These flows can be represented by the equations \cite{glotter2014simple}
\begin{align}
&  \frac{dC_A}{dt} = -k_a C_A + k_a \frac{k_H}{\delta_a }C_U + E(t) \text{ ,} \\
&\frac{dC_U}{dt} = k_a C_A - k_a \frac{k_H}{\delta_a}C_U -k_d C_U + \frac{k_d}{\delta_d} C_L \text{ ,} \\
& \frac{dC_L}{dt} = k_d  C_U - \frac{k_d}{\delta_d} C_L \text{ .}
\end{align}

However, this does not completely model the dynamics between the ocean and the atmosphere. We must take into account the dynamic ocean chemistry  as atmospheric \coo is dissolved in saltwater into the forms of bicarbonate ($HCO_3^-$) and carbonate ($CO_3^{2-}$). The dissociation of carbon into  its compounds occur with dissociation constants $k_1$ and $k_2$ (see Table \ref{tab:params}). This dynamic equilibrium is represented by \cite{glotter2014simple}
\begin{align}
    & CO_2 + H_2 O \overset{k_1}{ \rightleftharpoons}HCO^-_3 + H^+  \overset{k_2}{\rightleftharpoons} CO_3^{2-} + 2H^+ , \text{ where }\\
    & k_1 = \frac{[HCO_3^-][H^+]}{[CO_2(aq)]} \text{ and } k_2 = \frac{[CO_3^{2-}][H^+]}{[HCO_3^-]} \text{ .} \label{konstants}
\end{align}

Because \coo is a weak acid that reduces the ocean's capacity to absorb additional carbon dioxide, we must determine what the ocean's carbon storage capacity is at a given time. We use $\Lambda$ as this ``carbon storage factor", representing the ratio of the equilibrium of total dissolved inorganic carbon to \coo (aq). Because we have the dissociation constants $k_1$ and $k_2$, $\Lambda$ also can be described as a function of hydrogen ion concentrations at a given time, and we get the following \cite{glotter2014simple}:
\begin{align}
    &\Lambda(t) = \frac{[CO_2]+ [HCO_3^-] + [CO_3^{2-}]}{[CO_2]} = 1 + \frac{k_1}{[H^+]} + \frac{k_1  k_2}{[H^+]^2} \text{.}
\end{align}
 
Here, $\Lambda$ is the ratio of the equilibrium of sum of dissolved inorganic carbon (DIC) relative to  \coo (aq).To find $\Lambda$ at a given time, we must know the concentration of hydrogen ions. By definition, we have the total dissolved inorganic carbon as 

\begin{align}
    [DIC] = [\coo(aq)] + [HCO_3^-] + [CO_3^{2-}] \text{ .} \label{DIC}
\end{align}
After some simple algebraic manipulation of the equations for $k_1$ and $k_2$ on \eqref{konstants} and plugging into our $[DIC]$ on \eqref{DIC}, we get 
\begin{align}
    [DIC] = [\coo(aq)] \left( 1 + \frac{k_1}{[H^+]} + \frac{k_1  k_2}{[H^+]^2} \right) \text{ .}
\end{align}
Because the charges of the ions must be balanced, we can calculate the our $[H^+]$ concentration using the alkalinity ($Alk$) as 
\begin{align}
    Alk & = [HCO_3^-] + 2[CO^-_3] = [\coo(aq)] \left( \frac{k_1}{[H^+]} + \frac{2 k_1  k_2}{[H^+]^2} \right) \\
    &   = \frac{C_U}{\Lambda} \left( \frac{k_1}{[H^+]} + \frac{ k_1  k_2}{[H^+]^2} \right) \text{ .}
\end{align}
Therefore, we can find the concentration of hydrogen ions at a specific time by solving the quadratic of
\begin{align} 
    [H^+]^2 + [H^+]k_1 (1 - \frac{C_U}{Alk} + k_1 k_2 \left(1 - \frac{2 C_U}{Alk} \right) =0
\end{align} where we take the  positive root to be our hydrogen concentration.

The ability to store carbon decreases significantly as the pH increases. Our whole ocean carbon chemistry and flux can be represented with

\begin{align}
&  \frac{dC_A}{dt} = -k_a C_A + k_a \frac{k_H}{\delta_a \Lambda}C_U + E(t) \text{ ,}  \label{atmos-ocean-eq}\\
 &\frac{dC_U}{dt} = k_a C_A - k_a \frac{k_H}{\delta_a \Lambda}C_U -k_d C_U + \frac{k_d}{\delta_d} C_L \text{ ,} \\
& \frac{dC_L}{dt} = k_d  C_U - \frac{k_d}{\delta_d} C_L \text{ .}
\end{align}

\newpage

\subsection{Radiative Forcing \label{rf}}

Radiative forcing (RF) is calculated for each greenhouse gas using equations provided in the IPCC's assessment's fifth assessment report   \cite{ar5_ch5_sup8}. RF from \coo has a log-linear temperature response to excess carbon past pre-industrial conditions ($C_0$). \meth and \nit have interacting RF due to infrared band overlap \cite{myhre1998} \cite{ar5_ch5_sup8}. Formulas for RF (W m$^{-2}$) are as follows:

\begin{align}
  & \Delta N_\coo = \alpha \times \ln \left( \frac{C}{C_0} \right) \text{ where } \alpha = 5.35\\
  & \Delta N_\meth = \alpha \times (\sqrt{M} - \sqrt{M_0}) - (f(M,N_0) - f(M_0,N_0) , \text{ where } \alpha = 0.036\\
  & \Delta N_\nit = \alpha \times (\sqrt{N} - \sqrt{N_0}) - (f(M_0,N) - f(M_0,N_0) , \text{ where } \alpha = 0.12 \\
  & f(M,N) = 0.47 \times \ln [ 1+2.01\times 10^{-5} (MN)^{0.75} + 5.31 \times 10^{-15} M (MN)^{1.52} ] \\
\end{align}
$C$, $M$, and $N$ refer to current atmospheric  concentrations of \coo in ppm, \meth in ppb, \nit in ppb respectively. $C_0$, $M_0$ and $N_0$ refer to concentrations in 1750. The $\alpha$ term is the radiative forcing coefficient with units of $\frac{W}{m^2}$ and varies with each equation as indicated.

We also include differential equations for \meth and \nit  concentrations as a function of emissions and a fixed decay rate to reflect their atmospheric perturbation lifetimes as given by the IPCC \cite{ar5_ch5_sup8}. We assume \coo does not decay from the atmosphere in the same way, but cycles into various ocean and terrestrial stores as described by the carbon cycle equations. The changes in \meth and \nit are represented by
\begin{align}
    & \frac{d\meth}{dt} = E_{anthro} + E_{bio} + E_{pf} - \frac{\meth}{\nu_1} \\
    &\frac{d\nit}{dt} = E_{anthro}  - \frac{\nit}{\nu_2}. 
\end{align}

where $\nu_1$ is 12 years and $\nu_2$ is 114 years. These perturbation lifetimes are approximations, as the lifetime for \meth is varied and depends on atmospheric conditions. We assume these rates to be constant.


\newpage

\subsection{Temperature Response}

Our model uses a two-box temperature response model described by Pierrehumbert et al. \cite{ph2014}. This gives an accurate temperature response based on historical atmospheric concentrations of greenhouse gases and for projected emission streams (see \ref{past}).

We use $T_{mix}$ to describe the temperature perturbation of the upper, or mixed, layer of the ocean. This model uses the mixed ocean layer temperature perturbation as a proxy for Earth surface temperature perturbation because ocean temperatures largely dictate surface temperature. $T_{deep}$ represents the temperature perturbation of the deep ocean. The model demonstrates the heat transfer between the two layers. We use $\mu$ to represent the heat capacity of each layer. The model uses $\gamma$ for the heat transfer coefficient between the mixed and deep layers, and $\hat{\lambda}$ for the climate sensitivity parameter, both in units of $\frac{W}{m^2K}$.

Temperature changes are driven by the change in radiative forcing, $\Delta$N. The radiative forcing for each greenhouse gas is calculated and added together as a net $\Delta$N, giving the following:

\begin{align}
  & \mu_{mix} \frac{dT'_{mix}}{dt} = - \hat{\lambda}T'_{mix} - \gamma(T'_{mix}-T'_{deep}) + \Delta N(t) \\
  & \mu_{deep} \frac{dT'_{deep}}{dt} = \gamma (T'_{mix} - T'_{deep}) 
\end{align}

\newpage

\begin{table}[!h]
    \centering
    \begin{tabular}{c|r |l | l}
      \multicolumn{4}{c}{\textbf{Model Parameters}} \\
        \toprule
        \textbf{Parameter} & \textbf{Value}  & \textbf{Unit} & \textbf{Description}  \\
         \midrule
         \multicolumn{4}{c}{Terrestrial Carbon Parameters (Hartin et al (2016))} \\\hline \hline
         $f_{ds}$ & 0.60 & & fraction of detritus carbon that transfers to the soil \\\hline
         $f_{ld}$ & 0.01 &  & fraction land use carbon that enters detritus  \\\hline
         $f_{ls}$ & 0.89 & & fraction land use carbon that enters soil\\\hline
         $f_{lv}$ & 0.10 & & fraction land use carbon that enters vegetation \\\hline
         $f_{nd}$ & 0.60 & & fraction of NPP carbon that enters the detritus \\\hline
         $f_{ns}$ & 0.05 & & fraction of NPP carbon that enters the soil\\\hline
         $f_{nv}$ & 0.35 & & fraction of NPP carbon that enters the vegetation \\\hline
         $f_{rd}$ & 0.25 & & fraction of respiration carbon that enters the detritus\\\hline
         $f_{rs}$ & 0.02 & & fraction of respiration carbon that enters the soil \\\hline
         $f_{vd}$ & 0.034 & & fraction of vegetation carbon that enters the detritus \\\hline
         $f_{vs}$ & 0.001 & & fraction of vegetation carbon that enters the soil\\\hline
         $\beta$ & 0.36 & & carbon fertilization parameter \\\hline
         $Q_{10}$ & 2.45 & & $Q_{10}$ respiration factor \\\hline \hline
         
         \multicolumn{4}{c}{Oceanic Carbon Parameters  (Glotter et al (2014))} \\\hline \hline
         $k_1$ & $8.00 \times 10^{-7}$& mol/kg & disassociation  constant \\ \hline
         $k_2$ & $4.63 \times 10^{-10}$  & mol/kg & disassociation constant \\ \hline
        $k_a $ &  $ 0.2 $&  yrs $^ {-1}$ & turnover time for upper ocean \\ \hline
        $k_d $ &  $0.05 $ &  yrs $^ {-1}$ & turnover time for deep ocean \\ \hline
        $k_H$ & $1.23 \times 10^3$ &  & Henry's constant \\ \hline
        Alk  & $767.0 \star$ & GtC & alkalinity of the ocean \\ \hline
        $\delta_d$ & 50 & & ratio between upper and lower ocean \\ \hline
        $\delta_a$ & calculated & & ratio between upper ocean and atmosphere \\\hline
        $AM$ & $ 1.77 \times 10^{20}$ & moles & moles of the atmosphere \\ \hline
        $OM$ & $7.8 \times 10^{22}$ & moles & moles of the ocean \\ \hline \hline
        
        \multicolumn{4}{c}{Temperature Parameters  (Pierrehumbert et al (2014))} \\\hline \hline
        $\mu_{mix} $ &  $3.154 \times 10^8$ & J m$^{-2}$ K$^{-1}$ & heat capacity of mixed layer \\\hline
        $\mu_{deep} $ & $6.307\times 10^9$ & J m$^{-2}$     K$^{-1}$ & heat capacity of deep ocean \\\hline
        $\hat{\lambda}$ & $1.2$ & W m$^{-2}$ K$^{-1}$ & climate sensitivity parameter\\\hline
        $\gamma $ & $1.2 $ &  W m$^{-2}$ K$^{-1}$ &  heat transfer coefficient \\\hline \hline
        
        \multicolumn{4}{c}{Permafrost Parameters  (Kessler et al (2017))} \\ \hline \hline
        $C_{PF}$ &  $ 1035 $ & GtC  & total GtC in the permafrost \\ \hline
        $ \tau $ &  $ 70 $ & years & e-folding time for decomposition \\ \hline
        $\beta $ & $ 0.172 $ & & coefficient for permafrost melt rate \\ \hline
        $\eta$ & 0.40 &  & fraction of non-labile carbon  \\\hline
        $\omega$ & 0.023 & & fraction of carbon released as methane \\ \hline

    \end{tabular}
    \caption{Model parameters. Terrestrial and oceanic carbon cycle values are used to calculate carbon flux between the terrestrial stores, oceanic stores, and the atmosphere. Rows without units are unitless values. In our model, we convert all masses (GtC) to moles. Parameters $k_1$ and $k_2$ were converted to mole fraction by multiplying by 18/1000. ($\star$) We raised the alkalinity parameter by a factor 1.02 to better match pre-industrial CO2 uptake rates. This is still within the reported experimental range for ocean alkalinity. Temperature response parameters  come from Pierrehumbert et al. We also convert $\gamma$ and $\hat{\lambda}$ from watts to joules per year when running the model.}
    \label{tab:params}
\end{table}
\newpage

\subsection{Permafrost Feedback}

A permafrost feedback is included in our model, as it is one of the  major short term climate feedbacks. Figure \ref{Fig. 1} shows how a portion of carbon in the permafrost becomes vulnerable as the permafrost thaws. This now vulnerable pool is subject to decomposition by microbial activity and thus emitted into the atmosphere as \coo or \meth. A portion of the permafrost carbon is considered fixed or ``passive", as it is non-labile and will not be released during the timescales of our study. We make the assumption that the extent of the permafrost decreases linearly with increase of global mean temperature and that carbon content of the permafrost is spatially homogeneous \cite{kessler2017}.

We follow the logic of Kessler's model \cite{kessler2017} but adapted into differential equation form. The total carbon in the permafrost is  $ C_{PF}$.  $L_c$ and $L_m$ combine to form the actively releasing labile carbon pool, where $L_c$ is the proportion that is destined to be released as \coo, and $L_m$ is the proportion that is emitted as \meth. Together, they represent the labile portion of the pool of carbon that is now vulnerable due to a decrease in permafrost extent. $L_c$ and $L_m$ each have their own emission stream $E(t)$. These pools are assumed to be 0 when initializing the model at year 2010.  The proportion of carbon that is emitted as \coo  or \meth is determined by the proportion parameter $\omega$. 

We describe the intact frozen carbon in the permafrost as the total carbon pool at initial time $t_0$ times the proportion of the permafrost left ($PF_{extent}$):

\begin{align}
    C_{frozen} = C_{PF}(t_0) PF_{extent}(t) .
\end{align}

We describe the extent of permafrost at time $t$ as
\begin{align}
    PF_{extent}(t) = 1- \beta \times (T(t) - T(t_0)) \text{,}
\end{align}
where the permafrost extent decreases linearly with rise in global mean temperature ($T$) above an equilibrium temperature ($T(t_0)$)\cite{kessler2017}. In our model, we use the temperature at year $2010$ as $T(t_0)$, while Kessler uses the temperature at the year 2000 as $t_0$, so our model slightly underestimates melting. Additionally, the model does not allow $C_{PF}$ to increase by permafrost refreezing.

 As temperature increases, the extent of the permafrost decreases, and there is newly thawed soil that is now vulnerable to microbial decomposition. We represent this decomposition as exponential decay \cite{kessler2017}.

The loss of carbon from the labile pool is represented by the exponential decay term $-\frac{1}{\tau}$. The carbon that has decayed from the permafrost is then added into our emissions stream.  This is the same for the active labile carbon pool being emitted as methane, because it is the remaining proportion ($\omega$), and it shares the same decay rate $-\frac{1}{\tau}$. 

The Kessler equations adapted into differential equation form are represented as

\begin{align}
   & \frac{dC_{PF}}{dt} = C_{frozen}(t) - C_{PF} = \rho: \hspace{1em} \rho \leq 0  \label{cpf} \\
   &\frac{dL_C}{dt} = -\rho \times(1-\eta)\times(1-\omega) - \frac{1}{\tau} L_C  \\
   & \frac{dL_M}{dt} = -\rho \times (1 - \eta) \times \omega - \frac{1}{\tau} L_M \\
   & E_{PF \text{ }  C}(t) = \frac{1}{\tau} L_C \label{lc}\\
   & E_{PF \text{ }  M}(t) = \frac{1}{\tau} L_M \label{lm} \hspace{.25cm}\text{.} 
\end{align}

\subsubsection*{Summary of Atmospheric Carbon}

With the permafrost emissions added, we generalize our atmospheric carbon pool as follows:


\begin{align}
\frac{dC_A}{dt} =  E_{anthro} + E_{PF} + E_{LC} + F_{T} + F_{O} \text{,}
\end{align}

where $E_{anthro}$ is fossil fuel emissions, $E_{PF}$ is the emissions from the permafrost, $E_{LC}$ is land use emissions, $F_{T}$ is the flux of carbon between the terrestrial stores from net primary productivity and respiration ($-NPP + RH_{det,soil}$ see Equations \eqref{vegeq}, \eqref{detriuseq} and \eqref{soileq}) , and $F_{O}$ is the flux of carbon into the oceanic sinks ($ -k_a C_A + k_a \frac{k_H}{\delta_a \Lambda}C_U$ see Equations \eqref{atmos-ocean-eq}).

\subsection{Validation and Initial Conditions} 

We begin by burning in the model by imposing atmospheric \coo concentration at 1750 levels and running the model for a few thousand years to allow the levels of carbon in the ocean and terrestrial stores to stabilize. We then use these stabilized values as our initial conditions at 1750 for the vegetation, detritus, soil, upper ocean and lower ocean boxes.

\begin{figure}[!ht]
    \centering
    \includegraphics[width=\textwidth]{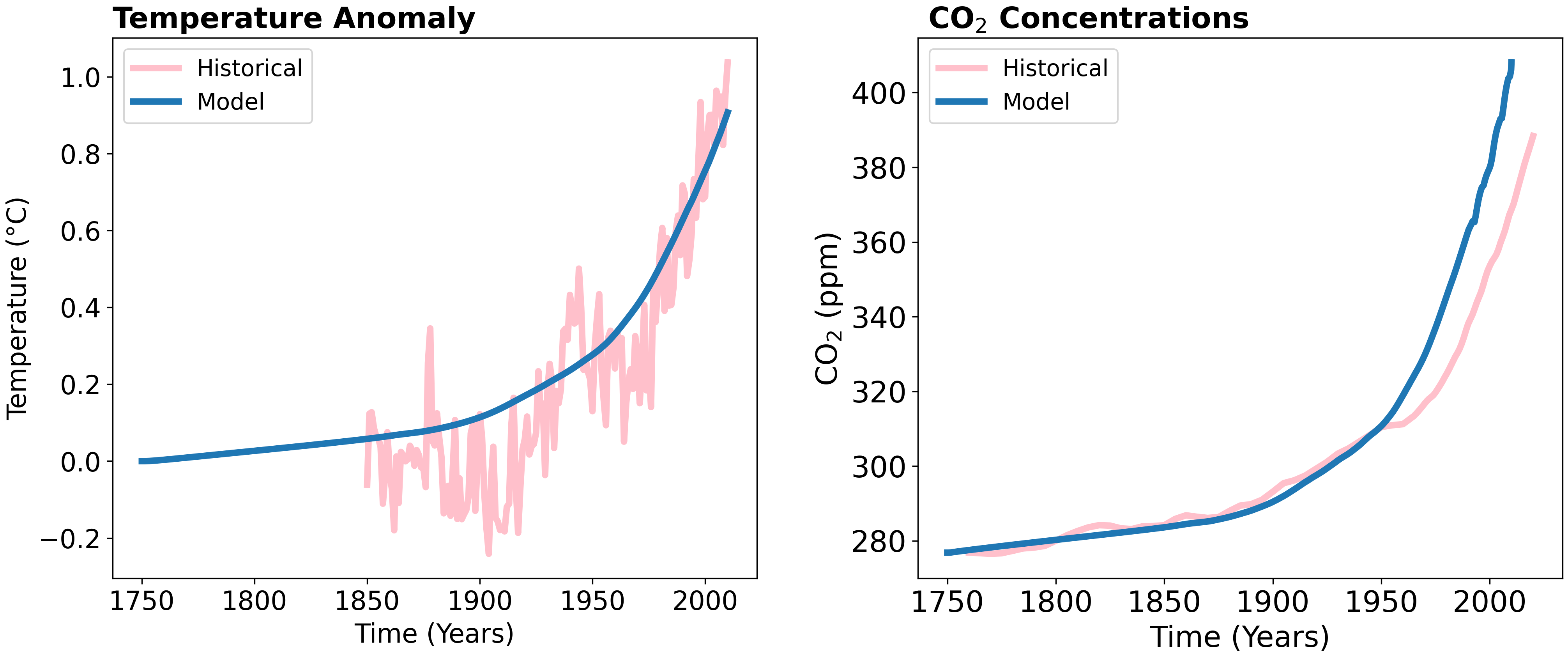}
    \caption{Validation of the model against historical temperatures and atmospheric \coo concentrations. (a) Historical temperature anomaly data is given in pink. The blue line gives the temperature anomaly as predicted by our model, driven by emissions data while solving for atmospheric carbon. (b) shows atmospheric carbon concentrations from historical data in pink, versus our solved carbon concentrations in blue. The overshoot in modeled \coo concentration past 1950 may be due to an unaccounted for increase biotic carbon uptake.}
    \label{past}
\end{figure}

We validated our model's temperature response and carbon cycle modules using historical data \cite{ar5_data}. Our initial conditions for 1750 are given in Table \ref{tab:Inits }. We drove the model from 1750 to 2010  using historical emissions data from the AR5 report \cite{ar5_data}, allowing our model to calculate atmospheric \coo concentrations based on its carbon cycle modules (Figure \ref{past} (b)). The model also calculated temperature from the radiative forcing from atmospheric \coo concentrations (Figure \ref{past} (a)), which shows a reasonably similar temperature result for 2010 \cite{hadleytemps}. 

To obtain initial conditions for the year 2010, we drove the rest of model using imposed historical \coo, \meth, and \nit atmospheric concentrations \cite{ar5_data} After so running the model to 2010, we took the carbon content of our upper ocean, lower ocean, vegetation, detritus, and soil boxes, along with the temperatures, as the initial conditions for projecting the model forwards.

\begin{table}[!h]
    \centering
    \begin{tabular}{c c || r | r | l r}
        \toprule
        \multicolumn{6}{c}{\textbf{Initial Conditions}} \\ \hline
        & &  \multicolumn{2}{c}{\textbf{Year}}  \\
        \textbf{Variable} & \textbf{Unit} & {1750} & {2010}  & \textbf{Description} & \textbf{Notes}\\  
            \midrule
        $T_{mix}$ &  $C \degree$ & 0 &  0.81 & surface temperature \\ \hline 
        $T_{deep}$ & $C \degree $ & 0 & 0.22 & lower ocean temperature\\ \hline
        $C_A$  & GtC  & 587.92 & 824.91  &  carbon in atmosphere & Glotter et al. 2014 \\ \hline
        $C_U$ &  GtC & 725.39 & 740.35 & carbon in upper ocean\\ \hline
        $C_L$ & GtC & 36263.18 & 36310.28 & carbon in lower ocean\\ \hline
        $C_v$ & GtC & 500 & 536.95  & carbon in vegetation  & Hartin et al. 2016 \\ \hline 
        $C_D$  & GtC & 55.29 & 59.73 & carbon in detritus \\ \hline
        $C_S$ & GtC & 1808.82 & 1767.11  & carbon in soil \\ \hline 
        \meth & ppb & N/A & 1798.0 & methane concentration & EEA 2019 \cite{EEA_2019}\\ \hline
        \nit & ppb & N/A & 323.7 & nitrous oxide concentration\\ \hline
        $C_{pf}$ & GtC  &  N/A & 1035  & carbon (frozen) in permafrost & Kessler et al. 2017 \\ \hline
        $L_C$  & GtC  & N/A & 0 & labile carbon \\ \hline
        $L_M$ & GtC & N/A & 0  & labile methane (as carbon) \\ \hline
    \end{tabular}
    \caption{Initial conditions for the year 2010, either found by driving the model with historical atmospheric concentration data or from other works (reference given in this case). These are the initial conditions used for our projections.}
    \label{tab:Inits }
\end{table}
\newpage

\subsection{Future Emission Scenarios}

To determine the impact of methane mitigation strategies on permafrost climate feedbacks, we projected our model forward using the AR5's extended RCP 2.6, 4.5, and 6.0 \coo and \nit emission streams \cite{rcpdata}. We combine these RCP \coo and \nit emission streams with our own calculated \meth emission stream based on different annual reduction plans. All our \meth emission streams start at 2010 with 330 Mt of \meth emitted that year and extend to 2300. To examine the impact of just the rate of phase-out, not magnitude, we take our initial value of 330 Mt and reduce it by a given percentage annually until it reaches the same ``floor" level of the given RCP. For example, RCP 2.6's projected methane emission stream falls from 330 Mt/year to 142 Mt/year by 2100, where it then remains until 2300. Therefore, we take 330 Mt, and reduce it by a given percentage annually until it reaches 142 Mt and then hold that level constant. We can then compare varying accelerated rates of phase-out to the given ``baseline" RCP scenario. For reference, the floor values for RCPs 4.5 and 6.0 are
266 Mt/year and 252 Mt/year, respectively.

To compare magnitude of phase-out, we take our projected methane emission streams to a lower final value than the RCP.

\newpage
\section{Results}

\begin{figure}[h!]{Temperature projections for all RCP Scenarios}
\centering
     \includegraphics[width=1.0\textwidth]{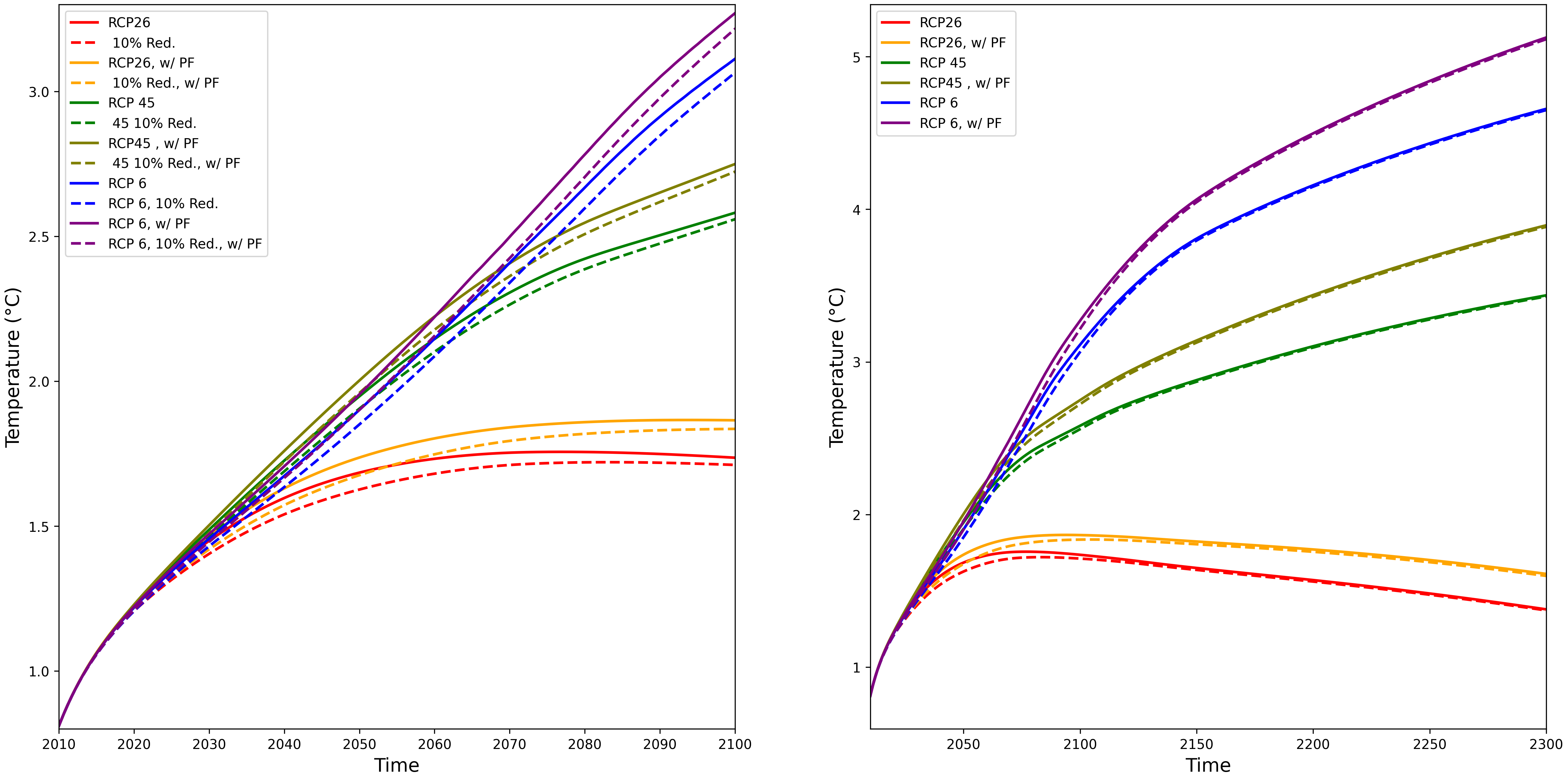}
    \caption{Temperature responses for RCP 2.6, 4.5, and 6 emission scenarios, with and without a permafrost feedback effect, (a) from 2010-2100 and (b) from 2010-2300. Dashed lines show a 10\% annual methane phase-out for the given scenario while solid lines show the baseline RCP scenario. Greatest differences in temperature between baseline RCP and 10\% annual phase-out are seen before 2100. From 2100-2300, the projections with and without the rapid 10\% methane reduction plans are largely the same, with or without the permafrost feedback in effect.}
    \label{TempResp}
\end{figure}

We began by generating temperature time series for all scenarios. First, to see the effect of the permafrost module, we ran our ``baseline" unaltered RCP emission scenarios from 2010 to 2300, and then ran the same scenarios again with the permafrost module in effect. We saw a significant change in temperatures with the permafrost module in effect, which can be seen in Figure \ref{TempResp}. Each RCP scenario has two main temperature projections, with the higher branch showing the permafrost feedback module in effect.

We also compared each RCP scenario with a rapid 10\% annual methane emission reduction plan, shown by the dashed lines in Figure \ref{TempResp}. Once the 10\% reduction plan reached the target methane emission rate of the given RCP, emissions were then held constant. For example, we compared the baseline RCP 2.6 with our RCP 2.6 with the 10\% annual methane phase-out; both reached the same final target of 142 Mt \meth emitted annually (but at different times), and remained at this level through 2300. This comparison was made with and without the permafrost feedback. A 10\% phase-out is considered very accelerated, and the target emission rate was reached within less than 5 years.

\newpage
\begin{table}
\centering
\begin{tabular}{l|c|c|c}
\hline
\multicolumn{4}{c}{\textbf{Temperature Anomaly from Permafrost}} \\ \hline
Year & RCP 2.6 & RCP 4.5 & RCP 6  \\
\hline
2100, without PF (\degree K) & 1.73 & 2.60 & 3.15 \\
2100, with PF (\degree K) & 1.86 & 2.77 & 3.31 \\
Difference (\degree K) & 0.13 & 0.17 & 0.16 \\ 
\% Difference & 7.5\% & 6.5\% & 5.1\% \\\hline
2300, without PF (\degree K) & 1.38 & 3.44 & 4.66 \\
2300, with PF (\degree K) & 1.61 &  3.9 & 5.13 \\
Difference (\degree K) & 0.23 & 0.46 & 0.47 \\
\% Difference & 16.7\% & 13.4\% & 10.1\% \\
\hline

\end{tabular}
\caption{Temperature perturbation at year 2100 and 2300, with and without the permafrost feedback effect under different RCP scenarios. Note that these results do not include accelerated methane phase-outs.}
\label{tab:PF2100table}
\end{table}

\begin{figure}[h!]{}
\centering
     \includegraphics[width=\textwidth]{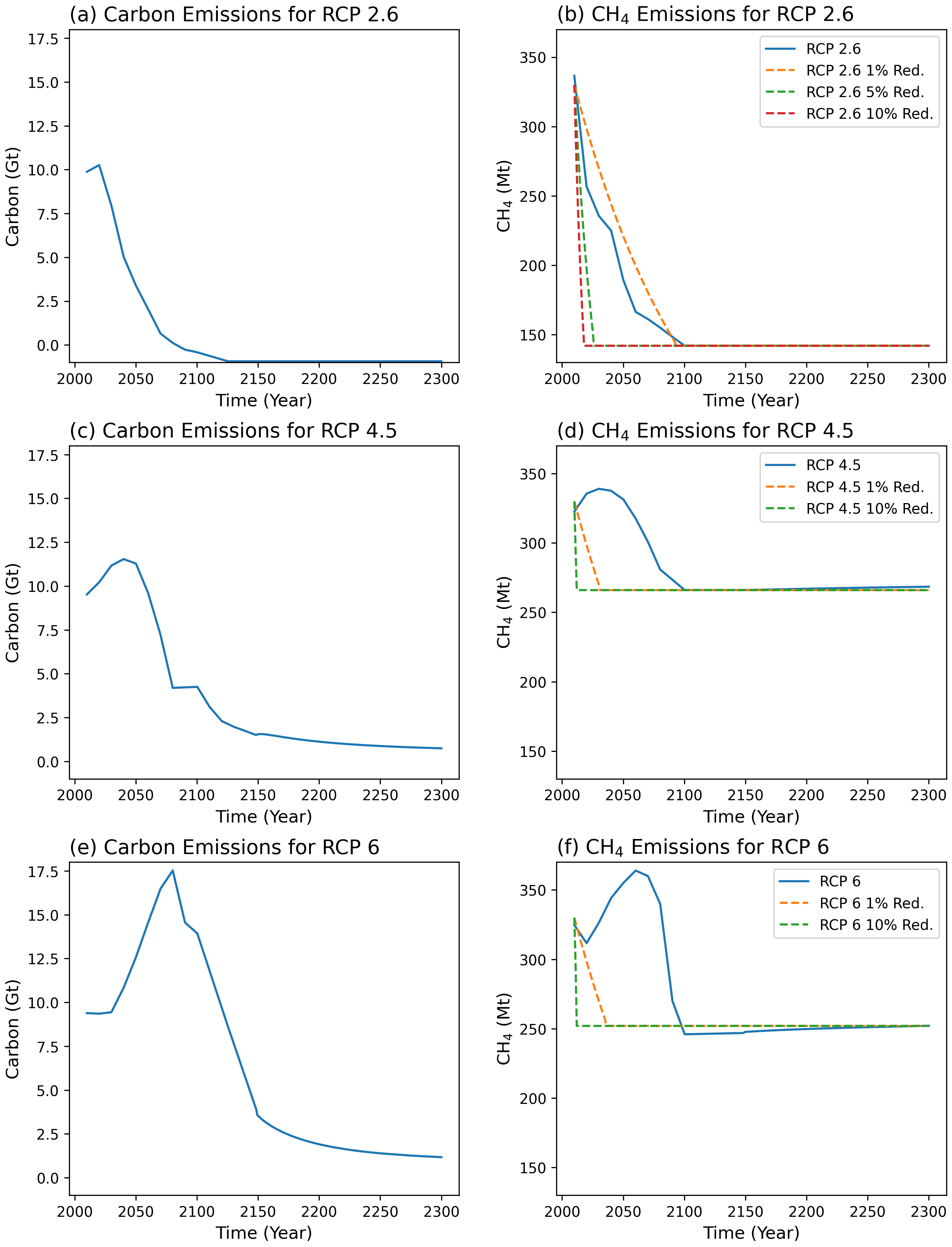}
    \caption{Emissions streams for all RCP scenarios and RCP scenarios with accelerated methane phase-out. Solid lines represent baseline RCP scenarios, while dashed lines represent our imposed methane mitigation scenarios.  Figures show total imposed emissions per year, not including permafrost feedback emissions. In Figures (a), (c), and (e), carbon emissions in gigatonnes are shown, as given by the RCP projection. In Figures (b), (d), and (f), \meth emissions in megatonnes are shown for the baseline RCP and our reduction scenarios.}
    \label{emissionsfigs}
\end{figure}

We compared our accelerated methane mitigation plans to the RCP's standard methane projections. For each RCP scenario, we used the IPCC's \coo emission projections, shown in gigatonnes of carbon in Figure \ref{emissionsfigs} (a), (c), and (e). We included each RCP's given \nit emission projections to accurately calculate methane's radiative forcing by taking into account the band overlap explained in section \ref{rf}.  We then compared the results of our methane mitigation plans with the baseline RCP scenarios. For RCP 2.6, we only considered 5\% and 10\% methane reduction rates, as a 1\% annual reduction was actually slower than the baseline RCP 2.6. For RCP 4.5 and 6.0, we used a 1\% and 10\% annual reduction rate, seen in Figure \ref{emissionsfigs} (b), (d), and (f). Figure \ref{emissionsfigs} shows these different emissions pathways.

\newpage

\newpage
\subsection{RCP 2.6}

\begin{figure}[h!]{RCP 2.6 Projections with 5\% and 10\% Annual Methane Reductions} 
\centering
     \includegraphics[width=1.0\textwidth]{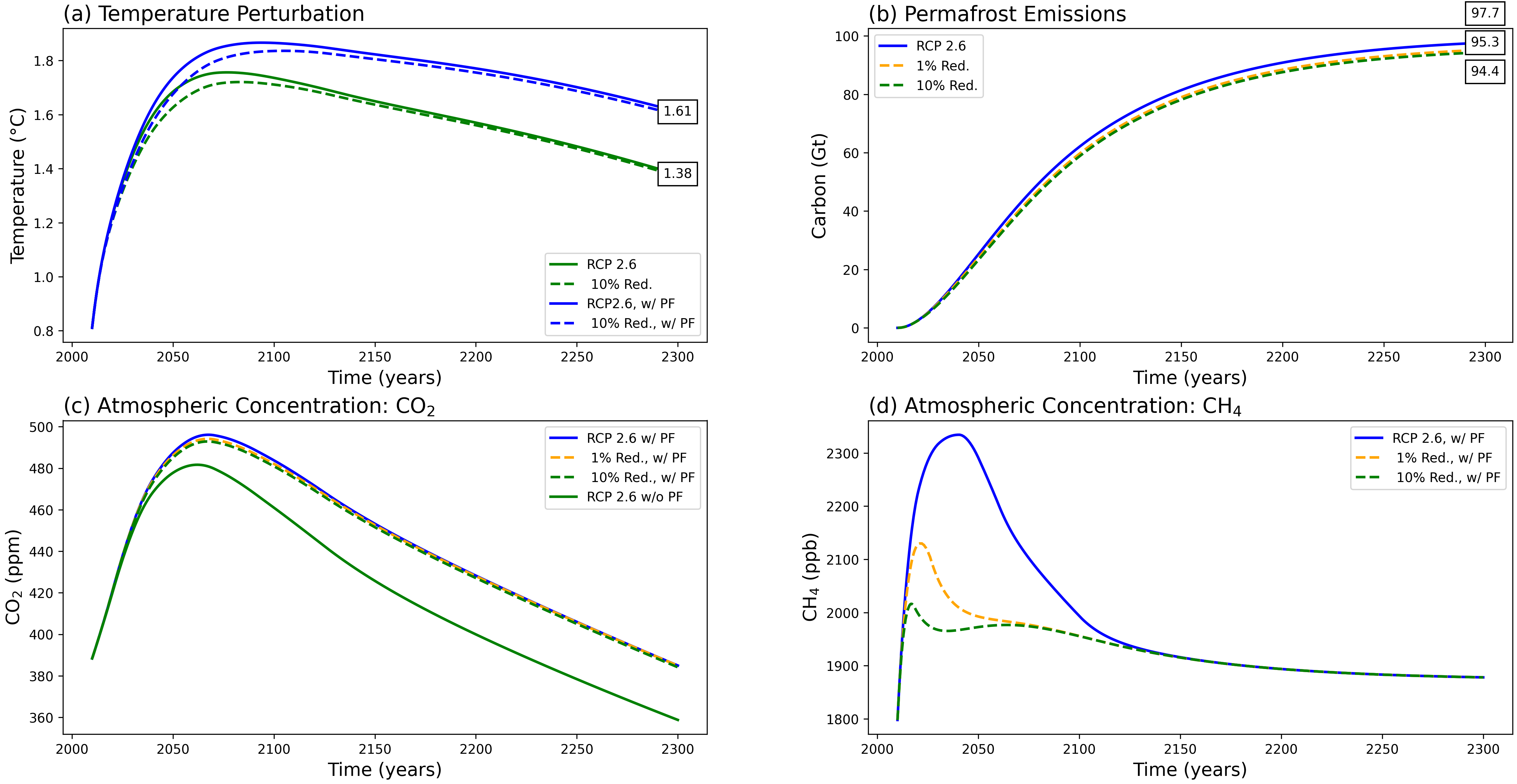}
    \caption{Results for 5\% and \%10 methane phase-out strategies using the RCP 2.6 \coo scenario. Dotted lines represent our imposed methane mitigation scenarios, while solid lines represent the ``baseline" RCP 2.6 scenario. We look at temperature perturbation at the year 2300 in (a), permafrost emissions at the year 2300 in (b), and atmospheric concentrations of \coo and \meth in (b) and (c). }
    \label{rcp26all}
\end{figure}

We ran the model under the RCP 2.6 scenario with and without the permafrost module to the year 2300 (see Figure \ref{rcp26all} (a)).  Without permafrost, the final temperature perturbation in 2300 reaches 1.38\degree C above pre-industrial levels. With the permafrost feedback, temperature reached 1.61\degree C, a difference of 0.23\degree C. Additionally,  the model gives 97.7 GtC emitted from the permafrost by 2300.

Using the RCP 2.6 scenario, we examined two highly accelerated methane reduction scenarios, where we reduced anthropogenic methane from 330 Mt in 2010 by 5\% or 10\% annually until we reached a constant emission rate of 142 Mt of \meth per year. The baseline RCP 2.6 scenario reaches a minimum or ``target" level of 142 Mt\meth per year at 2100, and remains at this level through 2300. Therefore, these accelerated mitigation scenarios only differed from RCP 2.6 in how long it took to reach that target. 

We found that the rate of methane mitigation had a negligible effect on temperature at 2300, but had a noticeable effect before 2100. The largest difference in temperature occurs at 2050, where there is a 0.06\degree C difference between the standard RCP 2.6 scenario compared to the 10\% phase-out (see SI \ref{tab:temptable}).

At the year 2300, cumulative 97.7 GtC was emitted from the permafrost under the baseline RCP 2.6, while 95.3 GtC was emitted under a 5\% accelerated methane reduction plan, and 94.4 GtC was emitted with a 10\% methane reduction plan. This may be because under RCP 2.6's aggressive \coo mitigation scenario, \meth emission mitigation has a relatively strong effect: We find that rapid methane mitigation has less effect on permafrost carbon release under other RCP scenarios.

Atmospheric concentrations of \coo are largely the same for all scenarios, and atmospheric concentrations for \meth reflect our differing reduction strategies, as well as slightly different permafrost emissions courses (see Figure \ref{rcp26all} (c) and (d). 

\newpage
\subsection{RCP 4.5}
We evaluated the impact of rate of mitigation under the RCP 4.5 emission scenario using the same basic procedures. 

First, we looked at temperature perturbation of the baseline RCP 4.5 scenario with and without a permafrost feedback effect. At 2300, the baseline RCP 4.5 resulted in a temperature anomaly 2300 of 3.44 \degree C, and with the permafrost module, 3.9\degree C (see table \ref{tab:PF2100table}). 277.8 GtC was emitted from the permafrost by year 2300.

\begin{figure}[h!]{RCP 4.5 Projections with 1\% and 10\% Annual Methane Reductions}
\centering
     \includegraphics[width=1.0\textwidth]{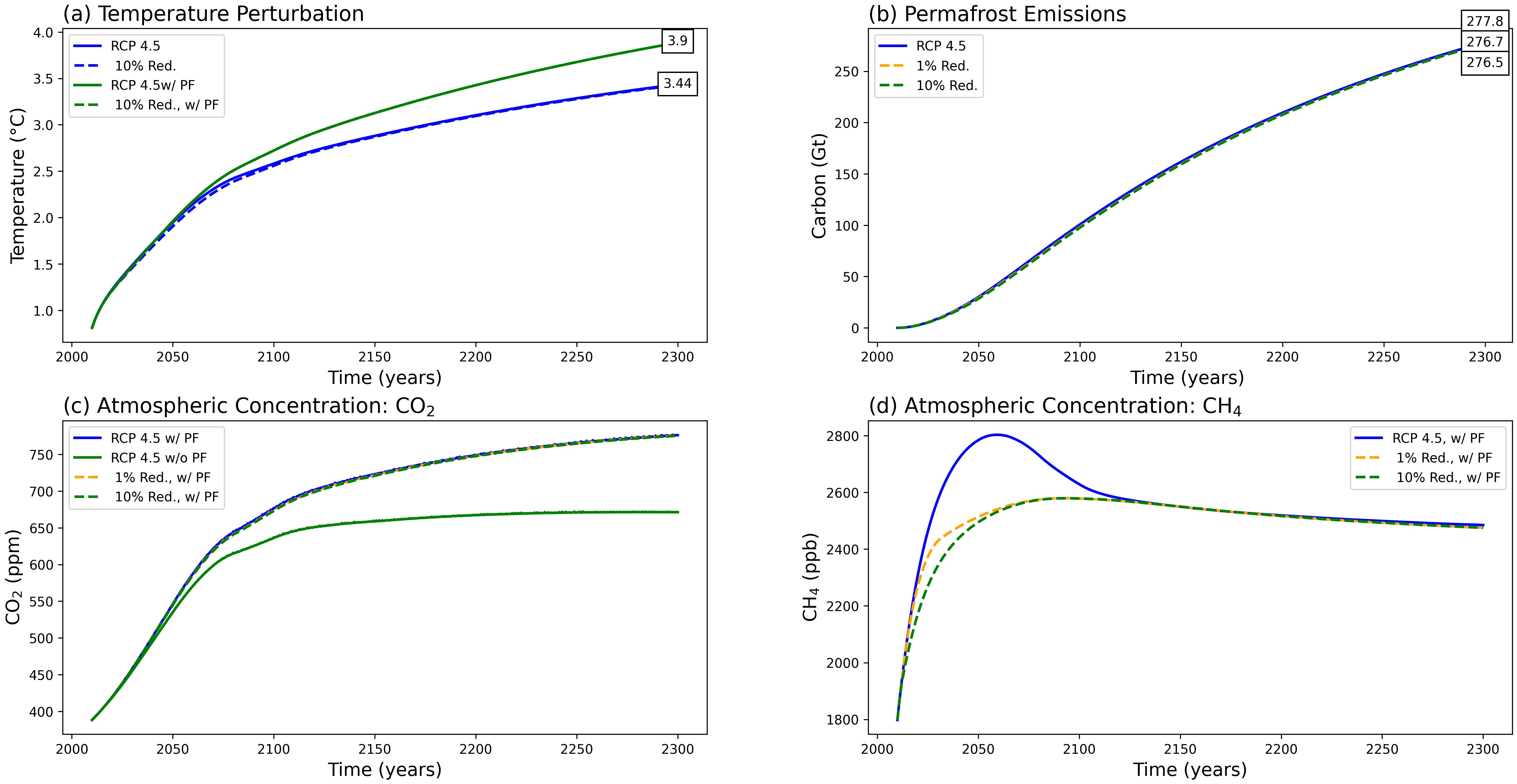}
    \caption{RCP 4.5 results for 1\% and \%10 reduction rates to annual methane emissions. (a) shows the final temperature perturbations with and without the permafrost feedback module in effect, with the dashed line showing the 10\% reduction scenario. (b) shows the permafrost emissions in gigatonnes of carbon for the baseline RCP scenario compared to the 1\% and 10\% reduction plans.(c) and (d) show atmospheric concentrations of \coo and \meth respectively over time.}
    \label{rcp45all}
\end{figure}

The target value of \meth emissions per year was set to 266 Mt\meth per year, consistent with the value the RCP 4.5 reaches at 2100 and remains constant at through 2300. We performed a 1\% or 10\% methane mitigation plan to reduce emissions from 330 to 266 Mt \meth per year. In Figure \ref{rcp45all} (d), the atmospheric concentrations of \meth are shown for the different scenarios. The mitigation plans reduced atmospheric methane concentrations compared to the baseline RCP before year 2100. After the year 2100, atmospheric concentrations remained the same between all scenarios, as expected since our reduction plans meet the same target value as the RCP 4.5.

The 1\% and 10\% phase-out schemes yield effectively the same final temperature perturbation, although there was a transient temperature difference before year 2100 (see Figure \ref{TempResp}). The temperature projections converged and were virtually identical by the year 2300 (see Table \ref{rcp45all}).

\newpage
\subsection{RCP 6}
\begin{figure}[h!]{RCP 6 Projections with 1\% and 10\% Annual Methane Reductions}
\centering
     \includegraphics[width=1.0\textwidth]{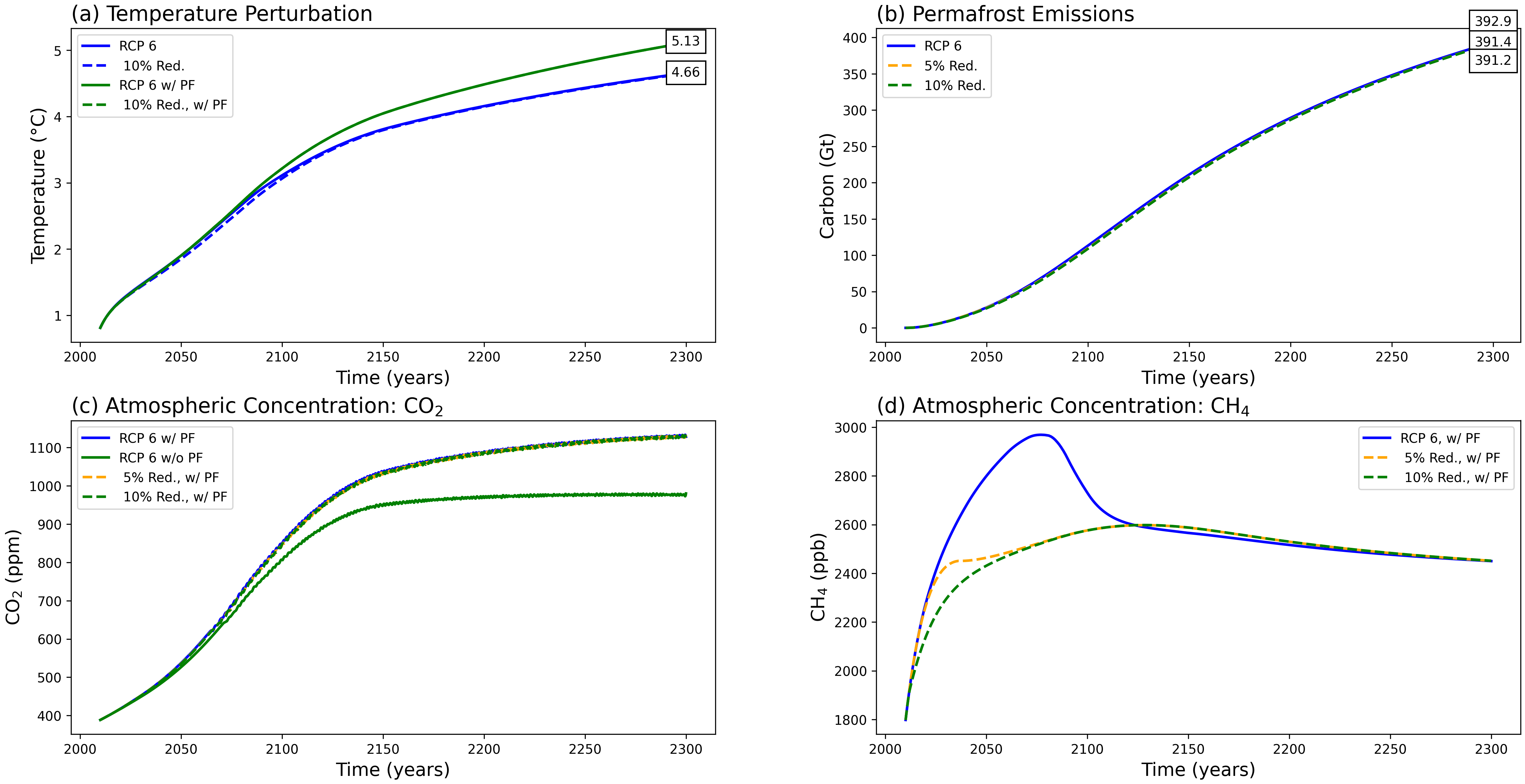}
    \caption{RCP 6 results for 1\% and \%10 methane reduction rates. (a) shows temperature perturbation at year 2300, with and without a permafrost feedback effect. Dashed lines denote an imposed methane reduction strategy, while solid lines represent the baseline RCP projection. (b) shows permafrost emissions in gigatonnes of carbon with a difference of 1.7 GtC at the year 2300 between the baseline RCP projection and the 10\% methane mitigation plan.(c) and (d) show atmospheric concentrations of \coo and \meth respectively.}
    \label{rcp6all}
\end{figure}

We evaluated methane mitigation plans compared with baseline RCP 6 projections (Figure \ref{rcp6all}). Temperature perturbation at the year 2300 without the permafrost module was 4.66 \degree C, and 5.13 \degree C with the permafrost module (see Table \ref{tab:PF2100table}). 392.9 GtC was emitted from the permafrost by the year 2300.

The target methane emission rate for RCP 6 was 252 Mt\meth yearly. Our methane mitigation scenarios went from 330 Mt\meth emitted in year 2010, reduced by 1\% or 10\% annually until reaching 252 Mt\meth and were then held constant for future years.

The temperature trajectories in Figure \ref{rcp6all} (a) were, in the long term, consistent between the baseline RCP projection and the imposed methane mitigation scenarios. There were significantly lower atmospheric concentrations of methane from the 2010-2100 under the mitigation strategies, however, this did not effect temperature projections or significantly impact permafrost emissions. Permafrost emissions were marginally lower (by 1.7 GtC) in the year 2100 from the 10\% mitigation scenario compared to the baseline RCP 6.

\newpage
\subsection{Final Magnitude Versus Rate of Mitigation}

Finally, we directly compared the effect of the final magnitude and rate of methane mitigation on final temperature perturbation at the year 2300 while including the permafrost feedback module. We used the RCP 4.5 scenario for this comparison. We ran the model with methane reduction rates between 0.1\% and 10\% to targets between 125 Mt \meth and 275 Mt \meth per year to obtain a temperature perturbation at the year 2300.  

Importantly, reduction rates below 0.5\% never reached the low end target of 125 Mt \meth emissions per year at 2300. If emissions were reduced by less than 0.5\% annually, there was not enough time from 2010 to 2300 to reduce emissions from 330 Mt \meth per year to 125 Mt \meth per year. For example, with a reduction rate of 0.1\%, there was still 246 Mt \meth being emitted at the year 2300, and with 0.25\% reduction rate, there was 160 Mt \meth emitted at 2300. This accounts for the difference in temperature perturbation at the year 2300 between the rates of 0.1 and 0.25\%  and the higher rates which converge to 3.7\degree C.

We saw a linear relationship between target methane annual emissions and final temperature perturbation.  Rates that were able to meet the target of 125 Mt \meth per year (1\%-10\% reductions annually) all show a strong cluster around 3.7\degree C final perturbation. The 0.5\% reduction scenario reached the target of 125 Mt\meth per year at year 2300, however, it reaches this goal much closer to year 2300 than the more accelerated plans. Thus, the spread around 3.7\degree C can be explained by this slower convergence to 125 Mt\meth. Overall, the final magnitude of methane emissions reductions proved to be more important than rate of reduction, provided the rate was fast enough to meet the target by year 2300.

\begin{figure}[h!]{}
\centering
     \includegraphics[width=1.0\textwidth]{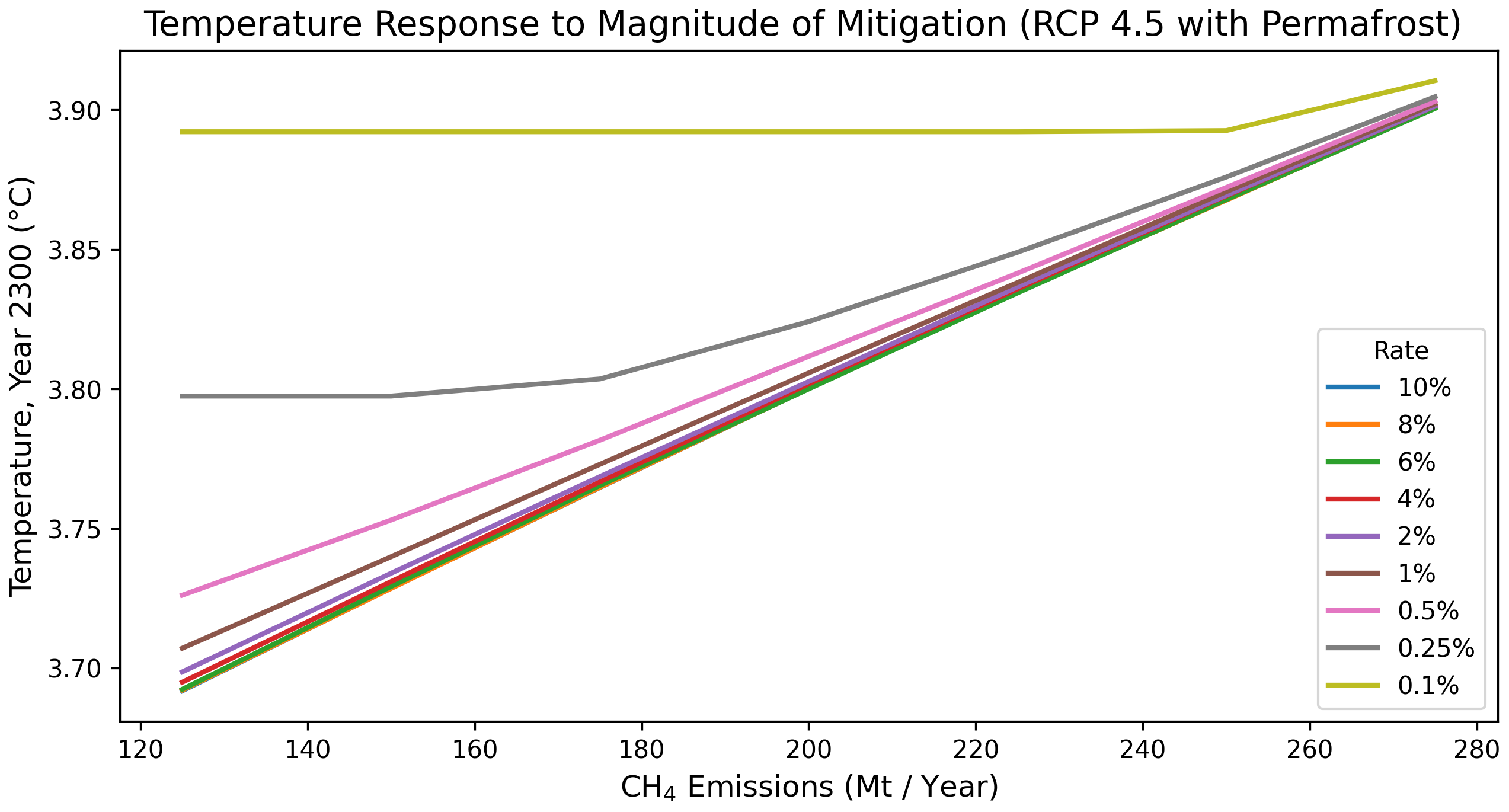}
    \caption{Final temperature perturbation in 2300, as a function of the rates of methane mitigation for different target levels, under RCP 4.5 with permafrost feedback. Importantly, with a very slow annual mitigation rate below 1\% each year (0.5\%, 0.25\%, 0.1\%) the final target of 125 Mt \meth per year was not reached. The magnitude of mitigation shows a strong relationship to final temperature, with a 0.2\degree Celsius difference in temperature between 120 Mt\meth and 275 Mt\meth emitted per year.}
    \label{mag}
\end{figure}

\begin{figure}[h!]{}
\centering
     \includegraphics[width=1.0\textwidth]{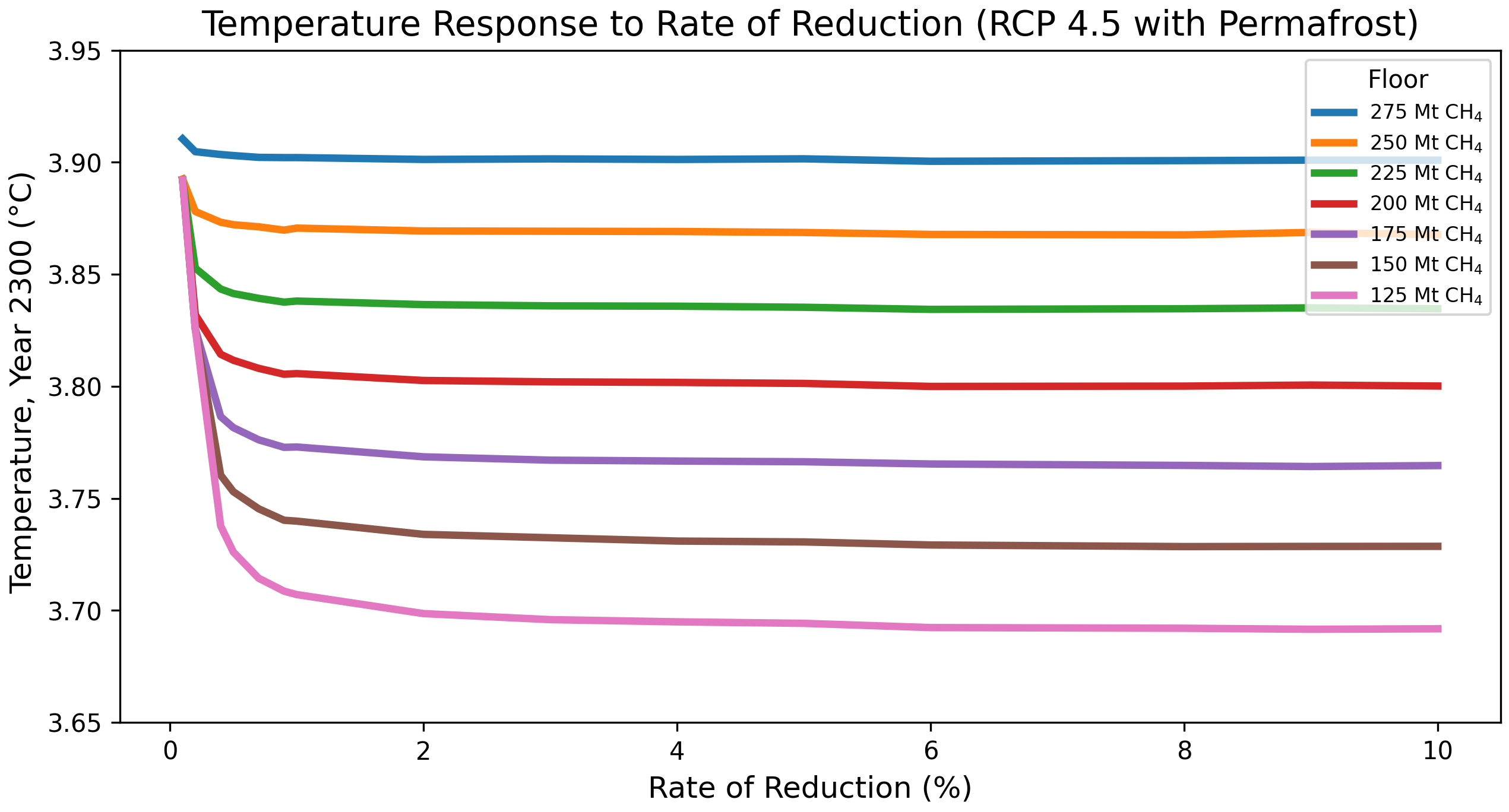}
    \caption{Temperature perturbation in the year 2300 as a function of the rate of methane mitigation, under RCP 4.5 with a permafrost feedback, and for different final emissions targets. For rates of reduction below 1\%, final mitigation targets are not met; therefore, temperatures are higher. For example, to reduce emissions to 125 Mt \meth /year, the rate of reduction must be over 1\% /year to achieve this target. When target emission rates are reached, temperature is determined by the final mitigation target.}
    \label{rate}
\end{figure}

Using RCP 4.5 with the permafrost feedback module, we then examined temperature perturbation in 2300 as a function of the rate of methane mitigation. When rates were below 0.5\%, methane mitigation targets were not reached. Between 1\% and 10\% annual reduction in emissions, the temperature response is largely the same, as shown by the horizontal lines in Figure \ref{rate}. We saw the same temperature perturbation across 1\%, 2\%, 4\%, 6\%, 8\%, and 10\% annual reduction for each final mitigation target (125, 150, 175, 200, 225, 250 and 275 Mt\meth/year). The final target mitigation level (assuming it was reached) determined the temperature perturbation. 

\newpage

\newpage

\begin{figure}[h!]
\centering
    \includegraphics[width=1\textwidth]{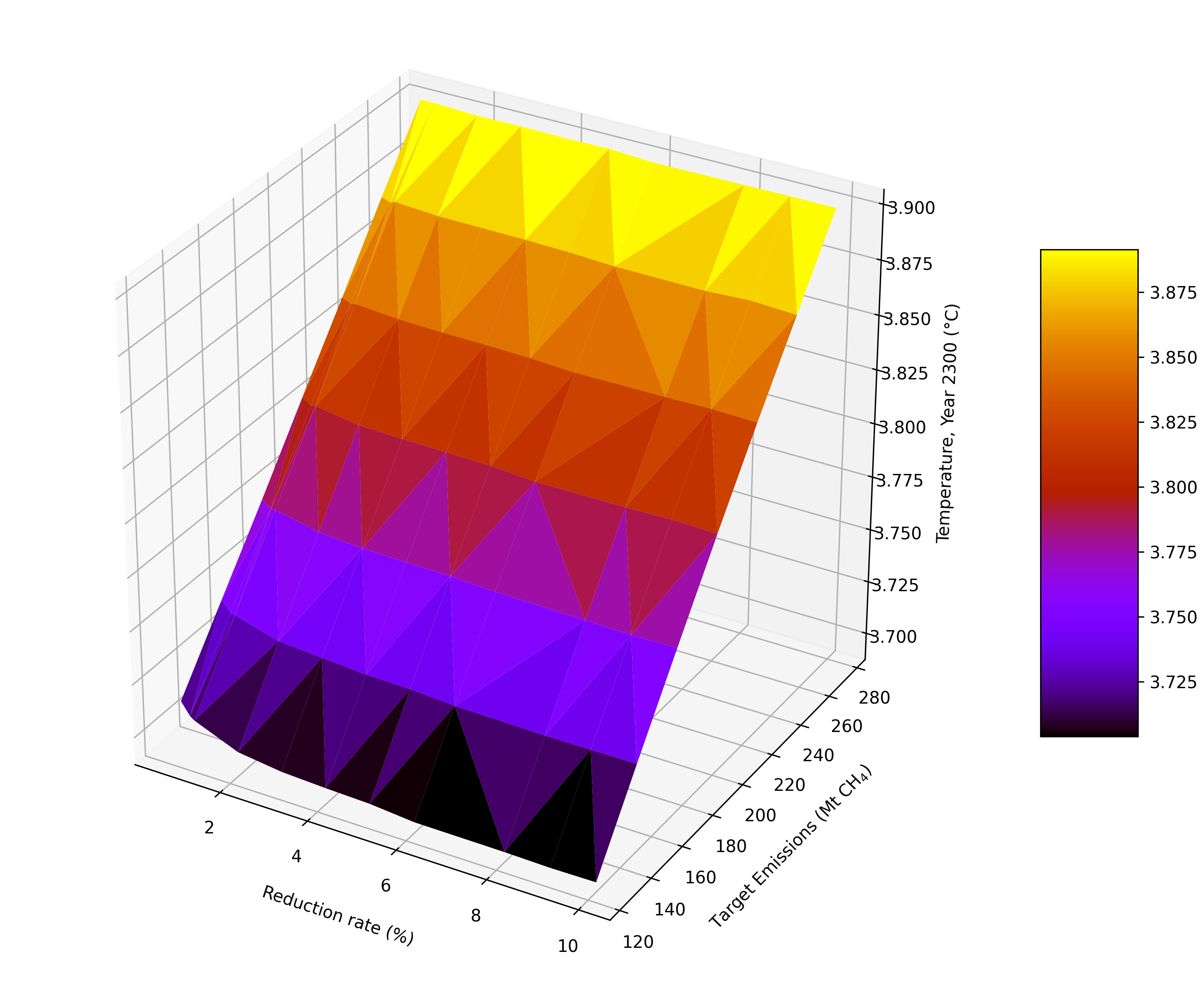}
    \caption{The effect of reduction rate and target emission level  on temperature perturbation at year 2300. Reduction rates vary between 0.5\% and 10\% and target emission levels vary between 125 and 275 Mt \meth. Time span is from 2010 to 2300, and the RCP 4.5 \coo emission scenario is used.}
    \label{quilt}
\end{figure}

In summary, we compared rate of methane reduction against target mitigation level and temperature (Figure \ref{quilt}). We used the same ranges as previously, excluding 0.1\% and 0.25\% reduction rates here as they could not reach the target mitigation level by 2300. For the rates shown between 0.5\% and 10\%, temperature perturbation was equal at year 2300. We found temperature perturbation responded highly linearly to the magnitude of mitigation, and not rate of reduction.

\newpage
\section{Discussion}

We developed and used a reduced complexity climate model to examine the long-term temperature perturbation as a function of methane emissions mitigation rates, in the presence of a permafrost feedback effect. We sought to clarify methane emission mitigation priorities. Mitigation of \coo emissions, with the goal of capping cumulative emissions, is critically important, as \coo has an extremely long atmospheric residence time. Thus, current emissions will affect climate for thousands of years, and a rapid emissions phase-out is urgent \cite{ph2014} \cite{ipcc15degrees}. Methane mitigation priorities, however, are much less clear, given this gas’ short atmospheric residence time. It is widely agreed that methane emissions should be reduced, but the timeline of how quickly this must be done is understudied \cite{dreyfus2022} \cite{nisbet2020}. Methane is atmospherically short lived and so the rate of emissions is the primary determinant of its climate effect.  However, its global warming potential 28 times that of \coo over a 100 year span, suggests  it may warrant an accelerated phase-out \cite{ar5_ch5_sup8}. Moreover, it is quite feasible that powerful short-term warming due to methane could accelerate permafrost thaw, and leading to large cumulative \coo emissions related to near-term methane emissions. Despite these concerns, our model predictions suggest that, even in presence of a strong permafrost feedback, while ultimate reductions in methane emissions are necessary, the \textit{rate} of mitigation is of minimal importance.

Looking at the permafrost feedback alone without methane mitigation, we found a significant difference in temperature perturbation. At year 2100, we saw a difference of 0.13-0.15\degree K (4.7-7.5\%) between all RCP scenarios, and at 2300, a difference of 0.23-0.47\degree K (10.1-16.7\%) (see Table \ref{tab:PF2100table}). We found the permafrost feedback to be a significant positive feedback that should be included in future climate modeling. 

We then generated accelerated methane mitigation scenarios to compare their long-term temperature perturbation with the IPCC's baseline RCP scenarios. We used the same methane emissions targets as the RCP scenario, and reduced anthropogenic methane emissions by 1\%, 5\% or 10\% annually until we reached that target. We found that mitigating methane at an accelerated rate did not prevent long term permafrost thaw or have a long term effect on temperature perturbation. Rather, we found that long temperature perturbation was largely determined by the magnitude of methane mitigation, not the rate of annual reduction. Provided the phase-out of methane was fast enough to reach the target level, the rate of annual reduction had a negligible effect.  Of course, to reach a lower target level, the rate of mitigation had to be sufficiently fast to hit the target on a relevant timescale; in the case of our model, that had to be before the year 2300.

While methane emissions must be mitigated, we found negligible long-term benefits to implementing a highly accelerated methane phase-out. When we reduced methane emissions to their target levels within 10 years with a 10\% annual reduction, we saw a slight difference in temperature this century (see Figure \ref{TempResp}). This, however, was transient cooling and \coo  ultimately determined the long term temperature course.

Our model recreates carbon fluxes and changes in radiative forcing while demonstrating a reasonable permafrost response. We aimed to capture the dynamics between anthropogenic emissions and permafrost thaw and decomposition, not conclusively predict future temperature anomalies. We included radiative forcing only from \coo, \meth, and \nit, which were able to adequately recreate surface temperature in the year 2010 where we start our projections. Black carbon, aerosols, halocarbons, ozone and other compounds could be included although they are not expected to make significant changes in the qualitative results, as historically, the cooling from aerosols has masked the warming from these other compounds \cite{dreyfus2022}. As \coo is mitigated, co-emitted aerosols will decline \cite{dreyfus2022}. However, with less cooling to mask other greenhouse gases, methane could end up being relatively more important.

The major source of uncertainty is permafrost modeling. There is great uncertainty regarding melting rates, microbial decomposition, and spatial heterogeneity \cite{deimling2012}. Our use of global mean temperature perturbation does not reflect regional temperature differences or spatial heterogeneity in carbon stores. We also make the assumption that permafrost thaw is a linear response to temperature increase.  Because there is great uncertainty, we chose a simple, high level representation of the permafrost.

A concern is the rate of microbial decay of permafrost carbon, or ``e-folding" time. We use 70 yr$^{-1}$, but it could be between 0-200 yr$^{-1}$ and may be variable \cite{kessler2017}. It is plausible that rapid methane mitigation is more important under faster decay rates. We, however, found our results insensitive to tao (see: Supplemental Information \ref{tab:taosens}). 



Another consideration may be the changing trends of methane-climate feedbacks  in the coming decades \cite{cheng2022}. There may be increasing levels of methanogenesis from wetlands and wildfires and a reduced methane atmospheric sink via hydroxyl radicals \cite{cheng2022}. These trends are not included in our model and are not expected to make a qualitative difference in model behavior. In our work, we use a constant value for atmospheric methane decay and a constant value for natural emissions. We also do not explicitly include the fossil methane, although upon testing this inclusion at a constant 25\% rate, we found negligible difference in our results (0.02\degree C difference at 2300 under RCP 4.5, see Supplemental Information \ref{tab:newcarbon}).


Our model gives warming attributable to permafrost within the range of other works. Macdougall et al. (2012) used a modified version of the Earth System Climate Model (ESCM) and found temperature difference attributable to permafrost thaw at 2100 to be 0.1-0.8\degree C, estimated at 0.27\degree C (consistent across different emission scenarios) \cite{macdougall2012}. Crichton et al. (2016) added a permafrost-carbon module to the Earth System Model of Intermediate Complexity(EMIC), CLIMBER-2, and  found increases of 10-40\% of the maximum temperature change \cite{crichton2016}. Burke et al. (2017) used the land surface model JULES (Joint UK Land Environment Simulator), along with ORCHIDEE-MICT (Organizing Carbon and Hydrology in Dynamic Ecosystems), modified to include permafrost carbon, coupled to the Integrated Model Of Global Effects of climatic aNomalies (IMOGEN), an intermediate-complexity climate and ocean carbon model. They found a difference of 0.2-12\% of the maximum temperature change \cite{burke2017}. Woodard et al. (2021) used a permafrost implementation in Hector v.2.3pf and found approximately 0.2\degree C or 4–15\% by 2100 across all RCP scenarios \cite{woodard2021}. 

In comparison, we found between 0.13 and 0.17\degree C (5.1-7.5\%) temperature perturbation attributable to the permafrost at 2100. This is on the low end but within the previously found range. We found a difference at 2300 (assumed maximum temperature change) of 0.23-0.47\degree C (10.1-16.7\%) which is on the low range of Crichton et al. but on the high range of Burke et al. \cite{burke2017}. 



Methane mitigation priorities are becoming clearer with recent research. By rapidly reducing anthropogenic methane to the greatest "economically feasible" extent, Ocko et al. (2021) saw a substantial near-term climate benefit at 2050 under RCP 8.5 \cite{ocko2021} and argue that because of this, rapid phaseout should be prioritized to minimize risk of overshooting climate boundaries and slow the rate of warming this century. However, by 2100 they found magnitude of mitigation was the most important factor. Although we focused on longer time scales and less extreme RCPs, we came to the same conclusions: fast mitigation does bend the temperature curve in the nearest term, with the largest benefit at 2050. However, by 2100, the fast mitigation and delayed mitigation pathways meet up to reach nearly the same temperature anomaly, and our inclusion of a large feedback does not change this result.

Research about methane mitigation priorities in relation to positive climate feedback effects is understudied. We find that the inclusion of the permafrost feedback does not indicate prioritizing rapid methane phaseout. Studies indicate that \coo is still the first priority, and no attention should be given to \meth mitigation at the expense of \coo mitigation \cite{mckeough2022} \cite{ph2014}. Our research agrees.

\section{Conclusions}
A benefit to an accelerated methane mitigation plan is that methane and \coo mitigation are coupled. As much of anthropogenic methane emissions are from fossil fuels (108-135 Mt\meth per year), \coo mitigation will also mean mitigating methane \cite{Jackson2020} \cite{rogelj2014}. As studies have stated, mitigating \coo immediately is of the utmost importance because of its long lasting climate effects, and methane mitigation is not a substitute \cite{ph2014} \cite{mckeough2022} \cite{rogelj2014}. Our research found that the extent of permafrost thaw was largely dependent on which RCP \coo emission projection we used and not the rate of methane mitigation, although methane mitigation must still occur. Bringing methane emission levels down to the lowest levels possible will play an important role in avoiding the climate boundary when done \emph{in addition} to mitigating \coo. By bringing down the RCP 4.5 target methane emissions down to RCP 2.6 levels, we saw an a 0.22\degree temperature difference at 2300.

We see marginal long-term benefit from rapid methane phase-out in relation to permafrost thaw. However, there is still a considerable short-term  benefit which may have effects on other sensitive feedbacks. Positive feedbacks, such as Amazon rainforest dieback, boreal forest loss, ice sheet melt, reduction in sea ice, etc., may amplify each other and have considerable effect on outcomes \cite{lenton2019}\cite{ritchie2021}. Inclusion of these other feedbacks and greater permafrost certainty would increase the confidence of the model. For this reason, we would still urge for methane mitigation at the fastest rate possible, but not at the expense of \coo mitigation.

\subsection*{Data Availability}
All data used in this work is cited and publicly available. No new primary data was created.

\subsection*{Declarations}
The authors declare that they have no known competing financial interests or personal relationships that could have appeared to influence the work reported in this paper.


\newpage

\section*{Supplemental Information}
\subsection*{Results}

Temperature anomaly values were taken at 2050, 2100, and 2300 with and without the permafrost module in effect.

\begin{table}[!h]
\centering
\begin{tabular}{l|c|c|c|c|c|c}
\hline
\multicolumn{7}{c}{\textbf{Temperature Anomalies}} \\ \hline
Scenario & 2050 & 2050 w/ PF & 2100 & 2100 w/PF & 2300 & 2300 w/ PF \\
\hline
2.6 & 1.70 & 1.75 &	1.73 & 1.86 & 1.38 & 1.61 \\
2.6, 5\% & 1.64 & 1.70 & 1.71 & 1.84 & 1.37 & 1.60 \\
2.6, 10\% & 1.64 & 1.69 & 1.71 & 1.84 & 1.37 & 1.60\\ \hline
        
4.5 & 1.99 & 2.05 & 2.60 & 2.77 & 3.44 & 3.90 \\
4.5, 1\%  & 1.95 & 2.01 & 2.58 & 2.75 & 3.43 & 3.89\\
4.5, 10\% & 1.94 & 2.00 & 2.58 & 2.74 & 3.43 & 3.89 \\ \hline
        
6 & 1.95 & 2.01 & 3.15 & 3.31 & 4.66 & 5.13 \\
6, 1\%  & 1.90 & 1.96 & 3.11 & 3.26 & 4.66 & 5.12\\
6, 10\% & 1.89 & 1.95 & 3.10 & 3.26 & 4.65 & 5.12 \\ 
\hline

\end{tabular}
\caption{Temperature perturbation at years 2050, 2100, and 2300, with and without permafrost feedback module, and per each annual methane emission reduction plan.}
    \label{tab:temptable}
\end{table}

\subsection*{Tao Parameter Sensitivity}

Tao is the parameter representing e-folding time of carbon in thawed permafrost soil. The sensitivity of this parameter was checked by using two extreme values within the range of possibility, 5 and 200 \cite{kessler2017}. Temperature anomaly at year 2300 was then compared for all RCP scenarios and results were as following:

\begin{table}[h!]
    \centering
    \begin{tabular}{l|c|c|c}
    \hline
        \multicolumn{4}{c}{\textbf{Temperature Anomaly at 2300}} \\ \hline
        & \multicolumn{3}{c}{Tao Values} \\ \hline
        Scenario & 5  & 70 & 200  \\ \hline
        RCP 2.6 & 1.65 & 1.61 & 1.54 \\
        RCP 4.5  & 3.98 & 3.9 & 3.74 \\
        RCP 6 & 5.13 & 5.23 & 5.05 \\
    \end{tabular}
    \caption{Temperature anomaly at year 2300 for varying tao values, representing different e-folding times for bacterial decomposition of permafrost carbon.}
    \label{tab:taosens}
\end{table}

Other model behavior did not change significantly. 

\newpage
\subsection*{Methane Oxidation}

Methane oxidation occurs when methane is exposed to ozone in the atmosphere and decomposes into carbon and water. Methane oxidation as a source of new carbon into the carbon cycle was left out of our model for simplicity. Here, we examine the inclusion of carbon from methane oxidation and show insignificant changes in temperature anomalies at year 2300. In our model, methane enters the atmosphere from permafrost emissions, anthropogenic emissions, and biogenic sources. We consider including methane oxidation from permafrost and fossil fuel emissions.  We determine that it does not significantly alter the behavior of our model.

\begin{table}[!h]
\centering
\begin{tabular}{l|c|c|c}
\hline
\multicolumn{4}{c}{\textbf{Temperature Anomaly under RCP 4.5 at 2300: Methane Oxidation Inclusion}} \\ \hline
Scenario & From Permafrost (1) & None (2)  & From Permafrost, Fossil Fuels (3) \\ \hline 
Baseline & 3.90 & 3.90 & 3.93\\ 
1\%  & 3.90 & 3.89 & 3.92 \\
10\%  & 3.89 & 3.89 & 3.92\\

\end{tabular}
\caption{Temperature anomaly at 2300 between differing inclusions of methane oxidation. We compare different inclusions between the baseline RCP 4.5, and 1\% and 10\% annual methane emission reduction scenarios. Column (1) shows inclusion of methane oxidation from permafrost methane emissions. Column (2) reflects no methane oxidation, as used in our paper. Column 4 shows inclusion of permafrost methane oxidation and 25\% of anthropogenic methane oxidation, an estimate used to reflect the fact that some sources of methane emissions are not from fossil fuels.}
    \label{tab:newcarbon}
\end{table}

\printbibliography

\end{document}